\documentclass[11pt]{article}

\newcommand{\calC}{{\cal C}}

\newcommand{\calL}{{\cal L}}

\newcommand{\bE}{{\bf E}}
\newcommand{\bP}{{\bf P}}

\newcommand{\tG}{{\tilde G}}

\newcommand{\tU}{{\tilde U}}
\newcommand{\tV}{{\tilde V}}

\newcommand{\tq}{{\tilde q}}
\newcommand{\tp}{{\tilde p}}
\newcommand{\tr}{{\tilde r}}
\newcommand{\tx}{{\tilde x}}

\newcommand{\trace}{{\rm Tr}}

\newtheorem{theorem}{Theorem}[section]
\newtheorem{prop}[theorem]{Proposition}
\newtheorem{lemma}[theorem]{Lemma}
\newtheorem{corollary}[theorem]{Corollary}
\newtheorem{remark}[theorem]{Remark}

\newcommand{\proof}{\noindent {\em Proof:~}}  

\newcommand{\nn}{\nonumber} 
\newcommand{\real}{{\bf R}}

\def\bd{\begin{displaymath}}
\def\ed{\end{displaymath}}

\def\eqref#1{(\ref{#1})} 
\def\qed{\hbox{\hskip 6pt\vrule width6pt height7pt depth1pt
    \hskip1pt}\bigskip}  
\def\to{\rightarrow}

\def\runinend{\enspace}
\def\ackname{Acknowledgement\runinend}%
\def\acknowledgements{\par\addvspace{17pt}\rmfamily
\def\ackname{Acknowledgements\runinend}%
\trivlist\if!\ackname!\item[]\else
\item[\hskip\labelsep
{\bf\ackname}]\fi}%

\begin{document} 
\bibliographystyle{plain} 

\hfill{} 
 
\thispagestyle{empty}

\begin{center}{\Large Exponential Convergence to Non-Equilibrium 
Stationary 
States in Classical Statistical Mechanics } 
\vspace{5mm}

Luc  Rey-Bellet\footnote{Email: lr7q@virginia.edu.}, 
Lawrence E. Thomas\footnote{Email: let@virginia.edu. Supported in part by 
NSF Grant 980139}\\    
\vspace{5mm}

\begin{center}
{\small\it Department of Mathematics, University of Virginia \\ 
Kerchof Hall, Charlottesville VA 22903, USA \\ }
\end{center}

\end{center}

\setcounter{page}{1} 
\begin{abstract} We continue the study of a model for heat conduction 
\cite{EPR1} consisting of a chain of non-linear oscillators coupled to
two Hamiltonian heat reservoirs at different temperatures. We
establish existence of a Liapunov function for the chain dynamics and
use it to show exponentially fast convergence of the dynamics to a
unique stationary state. Ingredients of the proof are the reduction of
the infinite dimensional dynamics to a finite-dimensional stochastic
process as well as a bound on the propagation of energy in chains of
anharmonic oscillators.

\end{abstract} 

\section{Introduction}\label{intro}

In its present state, non-equilibrium statistical mechanics is lacking
the firm theoretical foundations that equilibrium statistical
mechanics has. This is due, perhaps, to the extremely great variety of
physical phenomena that non-equilibrium statistical mechanics
describes. We will concentrate here on a system which is maintained,
by suitable forces, in a state far from equilibrium. In such an
idealization, the non-equilibrium phenomena, can be described by
stationary non-equilibrium states (SNS), which are the analog of
canonical or microcanonical states of equilibrium.

Recently many works have been devoted to the rigorous study of SNS.
Two main streams are emerging. In the first approach, for {\em open
systems}, a system is driven out of equilibrium by interacting with
several reservoirs at different temperatures.  In the second approach,
for {\em thermostated systems }, a system is driven out of equilibrium
by non-Hamiltonian forces and constrained to a compact energy surface
by Gaussian (or others) thermostats \cite{GC,Ru1}.  One should view
both approaches as two different idealizations of the same physical
situation, in the same spirit as the equivalence of ensembles in
equilibrium statistical mechanics.  But for the moment, the extent to
which both approaches are equivalent remains a largely open problem.

We consider here an open system, a model of heat conduction consisting
of a finite-dimensional classical Hamiltonian model, a one-dimensional
finite lattice of anharmonic oscillators (referred to as the chain),
coupled, at the boundaries only, to two reservoirs of classical
non-interacting phonons at positive and different temperatures. We
believe this model to be quite realistic, in particular it is
completely Hamiltonian and non-linear.

This model goes back (in the linear case) to \cite{FKM} (see also
\cite{RLL,SL}). First rigorous results for anharmonic models appear
\cite{EPR1} and further in \cite{EPR2,EH}.  Similar models in
classical and quantum mechanics have attracted attention in the last
few years, mostly for systems coupled to a single reservoir at zero
or positive temperature, i.e., for systems near thermal equilibrium
(see e.g. \cite{JP1,JP2,BFS,KKS,Ru2}. In our case, with two
reservoirs, no Gibbs Ansatz is available and in general, even the very
{\em existence} of a (non-equilibrium) stationary state is a
mathematically challenging question which requires a sufficiently deep
understanding of the dynamics. For the model at hand, conditions for
the {\em existence} of the SNS have been given in \cite{EPR1} and
generalized in \cite{EH}. The {\em uniqueness} of the SNS as well as
the strict positivity of {\em entropy production} (or heat flux) have
been proved in \cite{EPR2}. The leading asymptotics of the invariant
measure (for low temperatures) are studied in \cite{RT} and shown to be
described by a variational principle.

Under suitable assumptions on the chain interactions and its
interactions with the reservoirs, we establish the existence of a
Liapunov function for the chain dynamics. We then use this Liapunov
function to establish that the relaxation to the SNS occurs at an {\em
exponential} rate, and finally we prove that the system has a {\em
spectral gap} (using probabilistic techniques developed by Meyn and
Tweedie in \cite{MeTw}).

The Hamiltonian of the model has the form
\begin{equation}
H\,=\, H_B + H_S + H_I \,.
\label{h1}
\end{equation}
The two reservoirs of free phonons are described by wave equations in
$\real^d$ with Hamiltonian
\begin{eqnarray}
H_B\,&=&\, H(\varphi_L,\pi_L) +  H(\varphi_R,\pi_R) \,, \nn \\
H(\varphi,\pi)\,&=&\, \frac{1}{2} 
\int dx \,(|\nabla\varphi(x)|^2 + |\pi(x)|^2)\,, \nn
\end{eqnarray}
where $L$ and $R$ stand for the ``left'' and ``right'' reservoirs, 
respectively.
The Hamiltonian describing the chain of length $n$ is given by 
\begin{eqnarray}
H_S(p,q)\,=\, \sum_{i=1}^n \frac{p_i^2}{2} + V(q_1, \cdots, q_n)\,, \nn \\
V(q)\,=\, \sum_{i=1}^n U^{(1)}(q_i) + \sum_{i=1}^{n-1}
U^{(2)}(q_i-q_{i+1}) \,. \nn
\end{eqnarray}
where $(p_i,q_i) \in \real^d \times \real^d$ are the coordinates and
momenta of the $i^{th}$ particle of the chain. The phase space of the
chain is $\real^{2dn}$.  The interaction between the chain and the
reservoirs occurs at the boundaries only and is of dipole-type
\bd
H_I \,=\, q_1 \cdot \int dx\, \nabla \varphi_L(x)\rho_L(x)  + 
q_n \cdot \int dx \,\nabla \varphi_R(x)\rho_R(x)\,, 
\ed
where $\rho_L$ and $\rho_R$ are coupling functions (``charge
densities'') which we will assume spherically symmetric.

Our assumptions on the anharmonic lattice described by $H_S(p,q)$ are
the following:
\begin{itemize}
\item {\bf H1 Growth at infinity}: The potentials $U^{(1)}(x)$ and
$U^{(2)}(x)$ are $\calC^\infty$ and grow at infinity like
$\|x\|^{k_1}$ and $\|x\|^{k_2}$: There exist constants $C_i$, $D_i$,
$i=1,2$ such that
\begin{eqnarray}
\lim_{\lambda \rightarrow \infty}  \lambda^{-k_i}    U^{(i)}( \lambda x ) 
\,&=&\, a^{(i)} \|x\|^{k_i} 
\,, \label{g1} \\
\lim_{\lambda \rightarrow \infty}  \lambda^{-k_i +1} \nabla U^{(i)}( 
\lambda x ) 
\,&=&\, a^{(i)} k_i 
\|x\|^{k_i-2}  x \,, \label{g2} \\
\| \partial^2 U^{(i)}(x) \| \, &\le& \, 
( C_i + D_i V(x))^{1-\frac{2}{k_i}}  \,. \label{xcx} 
\end{eqnarray}
where $\| \cdot\|$ in Eq. \eqref{xcx} denotes some matrix-norm. 

Moreover we will assume that
\bd
k_2 \,\ge \, k_1 \, \ge \, 2\,,
\ed
so that, for large $\|x\|$ the interaction potential $U^{(2)}$ is
"stiffer" than the one-body potential $U^{(1)}$. It follows from
Eqs. \eqref{g1} and \eqref{g2} that the critical set of $V(q)$, i.e,
the set $\{ q\,:\, \nabla V(q)=0\}$ is a compact set.

\item{\bf H2 Non-degeneracy}: The coupling potential between nearest
neighbors $U^{(2)}$ is non-degenerate in the following sense.  For
$x\in {\bf R}^d$ and $m=1,2, \cdots$, let $A^{(m)}(x): {\bf R}^d
\rightarrow {\bf R}^{d^{m}}$ denote the linear maps given by
\bd
(A^{(m)}(x) v)_{l_1 l_2 \cdots l_{m}} \,=\, 
\sum_{l=1}^d \frac{\partial^{m+1}U^{(2)}}{\partial x^{(l_1)} 
  \cdots \partial x^{(l_m)} \partial x^{(l)}}(x) v_l \,.
\ed
We assume that for each $x \in {\bf R}^d$ there exists $m_0$ such that 
\bd
{\rm Rank} ( A^{(1)}(x), \cdots A^{(m_0)}(x)) = d \,.
\ed
\end{itemize}

The class of coupling functions $\rho_i$, $i\in \{L,R\}$ we can allow 
is relatively restrictive: 
\begin{itemize} 
\item{\bf H3 Rationality of the coupling}: Let ${\hat \rho}_i$ denote
the Fourier transform of $\rho_i$.  We assume that
\bd
|{\hat \rho}_i(k)|^2 \,=\, \frac{1}{Q_i(k^2)}\,,
\ed
where $Q_i$, $i\in \{L,R\}$ are polynomials with real coefficients and 
no roots on the real axis.  In particular, if $k_0$ is a root of $Q_i$, 
then so are $-k_0$, ${\overline k_0}$ and $-{\overline k_0}$.
\end{itemize}

Under these conditions we have the following result (a more detailed
and precise statement will be given in the next section).  Let
$F(p,q)$ be an observable on the phase space of the chain, for
example any function with at most polynomial growth (no smoothness is
required). We denote as $(p(t),q(t))$ the solution of the Hamiltonian
equation of motion with Hamiltonian \eqref{h1} and initial conditions
$(p,q)$.  Of course $(p(t),q(t))$ depends also on the variable of the
reservoirs, though only through their initial conditions
$(\pi_L,\varphi_L, \pi_R, \varphi_R)$.  We introduce the temperature
by making the assumption that the initial conditions of the reservoirs
are distributed according to thermal equilibrium at temperature $T_R$
and $T_L$ respectively and we denote $\langle \cdot \rangle_{LR}$ as the
corresponding average.

\begin{theorem} \label{main1}
Under Conditions ${\bf H1}-{\bf H3}$, there is a measure $\nu(dp,dq)$
with a smooth everywhere positive density such that the Law of Large
Numbers holds:
\bd
\lim_{T \rightarrow \infty} \frac{1}{T} \int_0^T F(p(t),q(t)) dt \,=\, 
\int F  \,d\nu
\ed
for almost all initial conditions $(\pi_L,\varphi_L, \pi_R,
\varphi_R)$ of the reservoirs and {\em for all} initial conditions
$(p,q)$ of the chain.  Moreover there exist a constant $r >1$ and a
function $C(p,q)$ with $\int C \,d\nu < \infty$ such that
\bd
\left|\langle F(p(t),q(t)) \rangle_{LR}  - \int F d\nu \right| \, \le \, 
C(p,q) 
r^{-t}\,. 
\ed  
for all initial conditions $(p,q)$.  That is, if we average over the
initial conditions of the reservoirs the convergence is {\em
exponential}.
\end{theorem}

Note that the ergodic properties stated in Theorem \ref{main1} hold
not only for $\nu$-almost every initial condition $(p,q)$, but in fact
for every $(p,q)$.

The existence of a (unique) stationary state was proved for (exactly
solvable) quadratic harmonic potentials $V(q)$ in \cite{SL}, for
$k_1=k_2=2$ (i.e., for potential which are quadratic at infinity) in
\cite{EPR1,EPR2} and generalized to the case $k_2 > k_1 \ge 2$ in
\cite{EH}.  What is really new here is that we prove that the
convergence occurs {\em exponentially fast} and we also weaken
slightly the conditions on the potential (in particular the case
$k_1=k_2$ is allowed and our condition {\bf H2} on $U^{(2)}$ is weaker
than the one used in \cite{EPR1,EPR2,EH}).  Our methods also differ
notably from those used in \cite{EPR1,EH}; in fact we reprove the
existence of the SNS (with a shorter and more constructive proof than
in \cite{EPR1,EH}) and, at the same time, we prove much stronger
ergodic properties.

We devote the rest of this section to a brief discussion of the
assumptions ${\bf H1}-{\bf H3}$.  Since the reservoirs are free
phonons gases and since we make a statistical assumption on the
initial condition of the reservoirs, one can integrate out the
variables of the reservoirs yielding random integro-differential
equations for the variable $(p,q)$. Our assumption ${\bf H3}$ of
rational coupling is, in effect, a Markovian assumption: with such
coupling one can eliminate the memory terms by adding a finite number
of auxiliary variables to obtain a system of Markovian {\em
stochastic} differential equations on the extended phase space
consisting of the dynamical variables $(p,q)$ together with the
auxiliary variables. The main (new) ingredient in our proof is then
the construction of a {\em Liapunov function} for the system, which
implies, using probabilistic methods developed in \cite{AM,Nu,MeTw},
the exponential convergence.

To explain the construction of a Liapunov function, note that the
dynamics of the chain in the bulk is simply Hamiltonian, while at the
boundaries the action of the reservoirs results into two distinct
forces. There are {\em dissipative} forces which correspond to the
fact that the energy of the chain dissipates into the reservoirs.  This
force is {\em independent } of the temperature. On the other hand
since the reservoirs are infinite and at positive temperatures, they
exert (random) forces at the boundaries of the chain and these forces
turn out to be proportional to the temperatures of the reservoirs.

The construction of the Liapunov function proceeds in two steps.  In a
first step we neglect completely the random force, only dissipation
acts.  This corresponds to dynamics at temperature zero, and one can
prove that the energy decreases and that the system relaxes to a
(local) equilibrium of the Hamiltonian $H(p,q)$.  We establish the
{\em rate} at which this relaxation takes place (at sufficiently high
energies). In the second step we consider the complete dynamics and we
show that for energies which are much higher than the temperatures of
the reservoirs, the random force is essentially negligible with
respect to the dissipation.  This means that except for
(exponentially) rare excursions the system spends most of its time in
a compact neighborhood of the equilibrium points.  On the other hand,
in this compact set, i.e., at energies of order of the temperatures of
the reservoirs, the dynamics is essentially determined by the
fluctuations and to prove exponential convergence to a SNS one has to
show that the fluctuations are such that every part of the phase space
is visited by the dynamics.  To summarize, we control the dynamics at
any temperature by the dynamics at zero temperature.

This allows one to understand the meaning of our assumptions on the
potential $V(q)$. If we suppose that the energy has an infinite number
of local minima tending to infinity, the zero temperature (long-time)
dynamics is {\em not } confined to a compact energy domain and our
argument fails.  With regard to the condition $k_2\ge k_1$ in {\bf H1} on
the exponents of the potentials, since the results of \cite{ST} and
the rigorous proofs of \cite{MA,Ba}, it is known that stable (in the
sense of Nekhoroshev) localized states exist in non-linear
lattices. Consider, for example, an infinite chain of oscillators
(without reservoirs). Numerically and in certain cases rigorously
\cite{MA}, one can show the existence of {\em breathers}, i.e., of
solutions which are spatially (exponentially) localized and
time-periodic. Although the breathers occurs both for $k_1 > k_2$ and
$k_2\ge k_1$ they behave differently at high energies. For $k_1 >
k_2$, the higher the energy, the more localized the breather get (hard
breathers), while for $k_2\ge k_1$, as the the energy gets bigger the
breather become less and less localized (soft breathers).  In fact a
key point of our analysis is to show that at high energy, if the
energy $E$ of the initial condition is localized away from the
boundary, then after a time of order one, the oscillators at the
boundaries carry at least an energy of order $E^{2/k_2}$ so that the
chain system energy can relax into the reservoirs.

Although we believe that the existence of a SNS probably may not
depend too much on these localization phenomena, the rate of
convergence to the SNS presumably does. Our approach of controlling
the dynamics by the the zero-temperature dynamics may not be adequate
if the condition {\bf H1} fails to hold and so more refined
estimates on the dynamics are needed to show that these
localized states might be in fact destroyed by the coupling to the
reservoirs.

As regards the organization of this paper, Sec. \ref{result} presents
the effective stochastic differential equations for the chain, a
discussion of allowable interactions between the reservoirs and the
chain and a concise statement, Theorem \ref{main}, of the exponential
convergence. In Sec. \ref{dissipation} we discuss the dissipative
deterministic system (corresponding to reservoirs at temperature $0$),
Theorem \ref{detdiss}, and then we show the extent to which the random
paths follow the deterministic ones, Proposition \ref{tracking}. We
give a lower bound on the random energy dissipation, Corollary
\ref{randiss}.  We then conclude Sec. \ref{dissipation} by providing
the Liapunov function, Theorem \ref{liapunov}, and bounds on the
exponential hitting times on (sufficiently large) compact sets,
Theorem \ref{hitting}. In Sec. \ref{controlandfeller} we prove the
random process has a smooth law and at most one ergodic component,
improving slightly results of \cite{EPR1,EPR2,EH}. Finally in
Sec. \ref{proof} we conclude the proof of Theorem \ref{main} by
invoking results of \cite{MeTw} on the ergodic theory of Markov
process.

\section{Effective Equations} \label{result}

We first give a precise description of the reservoirs and of their
coupling to the system and derive the stochastic equations which we
will study.  A free phonon gas is described by a linear wave equation
in $\real^d$, i.e., by the pair of real fields $\phi(x)=(\varphi(x),
\pi(x))$, $x\in \real^d$.  We define the norm $\|\phi\|$ by $
\|\phi\|^2 \,\equiv \, \int dx\, |\nabla \phi(x)|^2 + |\pi(x)|^2$ and
denote $\langle \cdot , \cdot \rangle$ the corresponding scalar
product.  The phase space of the reservoirs at finite energy is the
real Hilbert space of functions $\phi(x)$ such that the energy
$H_B(\phi)= \|\phi\|^2/2$ is finite and the equations of motion are
\bd
{\dot \phi}(t,x) \,=\,  \calL \phi(t,x)\,, \quad \calL \,=\, 
\left( \begin{array}{cc} 0 & 1 \\ -\Delta & 0 \end{array}\right)\,.
\ed 
In order to describe the coupling of the reservoir to the 
system, let us consider first a single confined particle in $\real^d$
with Hamiltonian $H_S(p,q) = p^2/2 + V(q)$. As the Hamiltonian for the
coupled system particle plus reservoirs, we have
\begin{eqnarray}
H(\phi,p,q) \,&=&\, \frac{1}{2}\|\phi\|^2 + p^2 + V(q) + q \cdot \int dx 
\, 
\nabla \varphi(x) \rho(x) \nonumber \\ 
\,&=&\, H_B(\phi) + H_S(p,q) + q \cdot \langle\phi , \alpha \rangle \,, \nn
\end{eqnarray} 
where $\rho(x)$ is a real rotation invariant function and 
$\alpha= (\alpha^{(1)}, \cdots, \alpha^{(d)})$ is, in Fourier space, given 
by
\bd
{\hat \alpha}^{(i)} \,=\, \left( 
\begin{array}{c} -ik^{(i)} {\hat \rho}(k) / k^2 \\ 0  \end{array} \right) 
\,.
\ed
We introduce the covariance matrix $C^{(ij)}(t) = \langle
\exp{(\calL t)}\alpha^{(i)}\,,\, \alpha^{(j)}\rangle$. A simple computation
shows that
\bd
C^{(ij)}(t)\,=\, \frac{1}{d}\delta_{ij} \int dk \, |\rho(k)|^2
e^{i|k|(t-s)}\,,
\ed 
and we define a coupling constant $\lambda$ by putting 
$\lambda^2 = C^{(ij)}(0)= \int dk  |\rho(k)|^2 $.
The equations of motion of the coupled system are
\begin{eqnarray}
{\dot q}(t) \,&=&\, p(t) \,, \nn \\
{\dot p}(t) \,&=&\, - \nabla V(q(t)) - \langle\phi, \alpha \rangle\,, \nn\\
{\dot \phi}(t,k)\,&=&\, \calL \left( \phi(t,k) + q(t)
\cdot \alpha(k) \right)\,. \label{pl}
\end{eqnarray}
With the change of variables 
$\psi(k) = \phi(k) + q \cdot \alpha(k)$, Eqs. \eqref{pl} become 
\begin{eqnarray} 
{\dot q}(t) \,&=&\, p(t)\,, \label{e1} \nn\\
{\dot p}(t) \,&=&\, - \nabla V_{\rm eff}(q(t)) - \langle \psi, \alpha 
\rangle 
\label{e2}\,, \nn \\
{\dot \psi}(t,k)\,&=&\, \calL  \psi(t,k) 
+ p(t) \cdot \alpha(k)\,,\label{e3} 
\end{eqnarray}
where
$V_{\rm eff}(q) = V(q) -\lambda^2q^2/2$.
Integrating the last of Eqs. \eqref{e3} with initial condition $\psi_0(k)$ one finds
\bd
\psi(t,k)\,=\, e^{\calL t}\psi_0(k) + \int_0^tds\, e^{\calL (t-s)} 
\alpha(k)\cdot p(s)\,.
\ed
and inserting into the second of Eqs. \eqref{e2} gives
\begin{eqnarray}
{\dot q}(t) \,&=&\, p(t)\,, \label{e4} \nn \\
{\dot p}(t) \,&=&\, - \nabla V_{\rm eff}(q(t)) - \int_0^t\,ds C(t-s) p(s) 
- 
\langle \psi_0, e^{-\calL t}  \alpha\rangle \label{e5}\,. 
\end{eqnarray}
If we now assume that, at time $t=0$, the reservoir is at temperature
$T$, then $\psi_0$ is distributed according to the Gaussian measure
with covariance $T\langle\cdot, \cdot\rangle$ and then
$\xi(t)\equiv\langle\psi_0, e^{-\calL t} \alpha\rangle$ is a
$d$-dimensional stationary Gaussian process with mean $0$ and
covariance $C(t-s)$. Note that the covariance itself appears in the
deterministic memory term on the r.h.s. of Eq.\eqref{e5}
(fluctuation-dissipation relation).

By assumption {\bf H3} there is a polynomial $p(u)$ which is a real
function of $iu$ and which has its roots in the lower half plane
such that
\bd
\int dk \, |\rho(k)|^2  e^{i|k|(t-s)}
\,=\, \int_{-\infty}^{\infty} du\, 
\frac{1}{|p(u)|^2} e^{iu(t-s)}\,.
\ed
Note that this is a Markovian assumption \cite{DyMk}: $\xi(t)$ is
Markovian in the sense that we have the identity
$p(-id/dt)\xi(t)={\dot \omega}(t)$, where ${\dot \omega}(t)$ is a
white noise, i.e, the joint motion of $d^m\xi(t)/dt^m$ , $0\le m \le
{\rm deg\,}{p}-1$ is a (Gaussian) Markov process. This assumption
together with the fluctuation-dissipation relation permits, by
extending the phase space with a finite number of variables, to
rewrite the integro-differential equations \eqref{e5} as a Markov
process. Note that $\xi(t)$ can be written as \cite{DyMk}
\bd
\xi(t) \,=\, \int_{-\infty}^{\infty} k(t-t') d\omega(t')\,, \quad 
k(t) \,=\, \int du\, e^{iu t} p(u)^{-1} \,.
\ed
with $k(t)=0$ for $t\le 0$.  For example if $p(u) \propto iu+\gamma$
then $C(t) = \lambda^2 e^{-\gamma|t|}$. Introducing the variable $r$
defined by
\bd
\lambda r(t) \,=\,\int_0^t ds\, C(t-s) p(s) + \int_{-\infty}^t k(t-t') 
d\omega(t')  \,,
\ed
then we obtain from Eqs.\eqref{e5} the set of Markovian differential
equations:
\begin{eqnarray}
{\dot q}(t) \,&=&\, p(t)\,, \label{e6} \nn \\
{\dot p}(t) \,&=&\, - \nabla V_{\rm eff}(q(t)) - \lambda r(t) 
\,,\label{e7} 
\nn \\
{ dr}(t) \,&=&\, (- \gamma r(t) + \lambda p(t))\,dt + 
(2T \gamma)^{1/2} d\omega (t) 
\label{e8} \,.
\end{eqnarray}
If $p(u) \propto (iu + \gamma + i\sigma)(iu + \gamma - i\sigma)$ then
$C(t)\,=\,
\lambda^2 \cos(\sigma t) e^{-\gamma|t|}$ and introducing the two auxiliary 
variables $r$ and $s$ defined by  
\begin{eqnarray}
\lambda r(t) \,&=&\, \lambda^2 \int_0^t ds\, \cos(\sigma(t-s)) 
e^{-\gamma|t-s|} p(s) \nn \\  
&& \quad + (T \lambda^2 \gamma )^{1/2}  \int_{-\infty}^t  
\cos(\sigma(t-s)) e^{-\gamma|t-s|} d\omega(s)\,, \nn \\
\lambda s(t) \,&=&\, \lambda^2 \int_0^t dt\, \sin(\sigma(t-s)) 
e^{-\gamma|t-s|} p(s) \nn \\ 
&& \quad + (T \lambda^2 \gamma)^{1/2} \int_{-\infty}^t dt\, 
\sin(\sigma(t-s)) e^{-\gamma|t-s|}d\omega(s) \,, \nn
\end{eqnarray}
we obtain then  the set of Markovian differential equations:
\begin{eqnarray}
{\dot q}(t) \,&=&\, p(t)\,, \label{e9} \nn \\
{\dot p}(t) \,&=&\, - \nabla V_{\rm eff}(q(t)) - \lambda r(t) 
\,,\label{e10} 
\nn \\
{ dr}(t) \,&=&\, (- \gamma r(t) -\sigma s(t)+ \lambda p(t))\,dt  
+ (2T\gamma)^{1/2} d\omega(t) \label{e11} \,, \nn\\
{\dot s}(t) \,&=&\,  - \gamma s(t) +\sigma r(t)   \label{e12} \,.
\end{eqnarray}
Obviously others similar set of equations can be derived for arbitrary
polynomial $p(u)$.

Another coupling which we could easily handle with our methods occurs
in the following limiting case, see \cite{FKM}. Formally one wants to
take $C(t)= \eta^2 \delta(t)$. Note that this corresponds to a
coupling function with $|\rho(k)|^2=1$ in which case $\lambda^2 =
\infty$. A possible limiting procedure consists in taking a sequence
of covariances tending to a delta function and at the same time
suitably rescaling the coupling (see \cite{FKM}).  In this case one
obtains the Langevin equations which serve as commonly-used model system with
reservoir in the physics literature,
\begin{eqnarray}
{\dot q}(t) \,&=&\, p(t)\,, \label{e13} \nn \\
dp(t) \,&=&\, (- \nabla V_{\rm eff}(q(t)) - \eta^2 p(t))\, dt +  
(2T\eta^2)^{1/2} d\omega(t)  \label{e14} \,.
\end{eqnarray}

The derivation of the effective equations for the chain is a
straightforward generalization of the above computations.  Our
techniques apply equally well to any of the couplings above. However,
for simplicity, we will only consider the case where the couplings to
both reservoirs satisfy $|\rho_i(k)|^2 \propto k^2 + \gamma^2$,
$i=L,R$. For notational simplicity we set $T_1=T_L$ and $T_n=T_R$, we
denote $r_1$ and $r_n$ as the two auxiliary variables and we will use
the notations $r=(r_1,r_n)$, and $x=(p,q,r)\in X = \real^{2d(n+1)}$.
In this case we obtain the set of Markovian stochastic differential
equations given by
\begin{eqnarray}
{\dot q_1} \,&=&\, p_1\,, \label{e15} \nonumber \\
{\dot p_1} \,&=&\, - \nabla_{q_1} V_{\rm eff}(q) - \lambda r_1 
\,,\label{e16} 
\nonumber \\
{ dr_1} \,&=&\, (- \gamma r_1 + \lambda p_1)\, dt + 
(2T_1 \gamma)^{1/2} d\omega_1\,, \label{e17} \nonumber \\
{\dot q_j} \,&=&\, p_j\,, \quad\quad \quad \quad\quad j=2,\dots,n-1\,, 
\label{e18} \nonumber \\
{\dot p_j} \,&=&\, - \nabla_{q_j} V_{\rm eff}(q) \,\,\quad j=2,\dots,n-1 \,, 
\label{e19} \nonumber \\
{\dot q_n} \,&=&\, p_n \,, \label{e20}\nonumber  \\
{\dot p_n} \,&=&\, - \nabla_{q_n} V_{\rm eff}(q) - \lambda r_n 
\,,\label{e21} 
\nonumber \\
{ dr_n} \,&=&\, (- \gamma r_n + \lambda p_n)\,dt + 
(2T_n \gamma)^{1/2} d\omega_n \,,\label{e22}
\end{eqnarray}
where $V_{\rm eff}(q) = V(q) - \lambda^2q_1^2/2 - \lambda^2q_n^2/2$. From 
now 
on, 
for notational simplicity  we will suppress the index ``eff'' and consider 
$V=V_{\rm eff}$ as our potential energy. 

It will be useful to introduce the following notation. We define the
linear maps $\Lambda : \real^{dn} \to \real^{2d} $ by $\Lambda (x_1,
\ldots, x_n)= (\lambda x_1, \lambda x_n)$ and $T : \real^{2d} \to
\real^{2d} $ by $T(x, y) = (T_1 x, T_n y)$. With this we can rewrite
Eq.\eqref{e22} in the compact form
\begin{eqnarray}
{\dot q} \,&=&\, p \,,\nn \\
{\dot p} \,&=&\, -\nabla_q V - \Lambda^T r \,, \nn \\
{  dr} \,&=&\, (-\gamma r + \Lambda p)\,dt + (2\gamma T)^{1/2} d\omega \,. 
\label{e23}
\end{eqnarray}
The solution $x(t)$ of Eq.\eqref{e23} is a Markov process. We
denote $T^t$ as the associated semigroup,
\bd
T^tf(x) \,=\, \bE_x [ f(x(t)]\,,
\ed
with generator
\begin{equation}\label{generator}
L\,=\, \gamma \left( \nabla_r T \nabla_r - r \nabla_r\right) + 
\left( \Lambda p \nabla_r - r \Lambda \nabla_p \right) + 
\left( p \nabla_q - (\nabla_q V(q)) \nabla_p\right)\,,
\end{equation}
and $P_t(x,dy)$ as the transition probability of the Markov process
$x(t)$.  There is a natural energy function which is associated to
Eq.\eqref{e23}, given by
\bd
G(p,q,r)\,=\, \frac{r^2}{2} + H(p,q)\,.
\ed
A straightforward computation shows that in the special case $T_1=T_n=T$
\bd
Z^{-1} e^{- G(p,q,r)/T} 
\ed
is an invariant measure for the Markov process $x(t)$. 

Given a function $W: X \rightarrow {\bf R}$ satisfying $W\ge 1$ we
consider the following weighted total variation norm $\| \cdot \|_W$
given by
\begin{equation}\label{wvar}
\| \pi \|_W \,=\, \sup_{|f| \le W} |\int f d\pi|\,,
\end{equation}
for any (signed) measure $\pi$. 
We introduce norms $\|\cdot\|_\theta$ and Banach spaces 
$L^\infty_\theta(X)$ given by 
\begin{equation}\label{ltheta}
\|f\|_\theta \,=\, \sup_{x\in X} \frac{|f(x)|}{e^{\theta G(x)}}\,, \quad
L^\infty_\theta(X)\,=\, \{ f \,:\, \|f\|_\theta < \infty \}\,,
\end{equation} 
and write $\|K\|_\theta$ for the norm of an operator $K: L^\infty_\theta(X) 
\rightarrow L^\infty_\theta(X)$. 

Theorem \ref{main1} is a direct consequence of the following result

\begin{theorem} \label{main}
Assume that {\bf H1} and {\bf H2} hold. The Markov process $x(t)$ which solves 
\eqref{e23} has smooth transition probability densities, 
$P_t(x,dy) = p_t(x,y)dy$, with $p_t(x,y) \in 
\calC^\infty((0,\infty) \times X \times X )$. The Markov process $x(t)$ 
has a unique  invariant measure $\mu$, and $\mu$ has a $\calC^\infty$ everywhere 
positive density. 
For any $\theta$ with $0 < \theta < (\max\{T_1,T_n\})^{-1}$ there exist 
constants 
$r=r(\theta) > 1$ and $R=R(\theta) <\infty$  such that 
\begin{equation} \label{s01}
\| P_t(x,\cdot) - \mu \|_{\exp{(\theta G})}  \, \le \, R r^{-t} 
\exp{(\theta G(x))} \,,
\end{equation}
for all $x \in X$, (exponential convergence to the SNS) 
or equivalently
\bd
\|T^t - \mu\|_\theta \le  R r^{-t}\,,
\ed
(spectral gap).
Furthermore for all functions $f$, $g$ 
with $f^2$, $g^2 \in L^\infty_\theta(X)$ and all $t>0$ we have 
\bd 
\left| \int g T^tf \,d\mu  - \int f \,d\mu \int g \,d\mu \right|  \,\le 
\, R r^{-t} \|f^2\|_\theta^{1/2}
\|g^2\|_\theta^{1/2}\,,
\ed
(exponential decay of correlations in the SNS). 
\end{theorem}

The convergence in the weighted variation norm, Eq. \eqref{s01}, implies 
that the Law of large Numbers holds \cite{Ha,MeTw}. 

\begin{corollary} Under assumptions {\bf H1} and {\bf H2} $x(t)$ satisfies the 
Law of Large Numbers: For all initial conditions $x \in X$ and all 
$f \in L^1(X,d\mu)$ 
\bd
\lim_{T\to \infty} \frac{1}{T} \int_0^T f(x(t))\, dt \,=\, \int f\, d\mu 
\ed
almost surely. 
\end{corollary}

The convergence of the transition probabilities as given in
\eqref{s01} is shown in \cite{MeTw} to follow from the following
properties:

\begin{itemize}
\item {\bf Strong Feller property} The diffusion process is strong
Feller, i.e., the semigroup $T^t$ maps bounded measurable functions
into continuous functions.
\end{itemize}
This is a consequence of the hypoellipticity of the diffusion $x(t)$,
which follows from Condition {\bf H2}, see Section
\ref{controlandfeller}.

\begin{itemize}
\item {\bf Small-time open set accessibility}. 
For all $t>0$, all $x\in X$ and all open set $A\subset X$ we have 
$P_t(x,A) > 0$.  
\end{itemize}
This means that the Markov process is ``strongly aperiodic''. In
particular, combined with the strong Feller property it implies
uniqueness of the invariant measure. This property is discussed in Section
\ref{controlandfeller} using the support theorem of \cite{SV} and explicit 
computations. This generalizes (slightly) the result obtained in
\cite{EPR2}.

\begin{itemize}
\item{ \bf Liapunov function and hitting times} Fix $s>0$
arbitrary. Set $W =\exp(\theta G)$ and choose $\theta$ with $0 <
\theta < (\max\{T_1,T_n\})^{-1}$. Then $W$ is a Liapunov function for
the Markov chain $\{x(ns)\}_{n\ge 0}$: $W > 1 $, $W$ has compact level
sets and there is a compact set $U$, (depending on $s$ and $\theta$)
and constants $\kappa < 1$ and $b < \infty$, (both depending on $U$,
$s$ and $\theta$) such that
\begin{equation}  \label{ll}
T^s W(x) \,\le \, \kappa W(x) + b {\bf 1}_U(x)\,,
\end{equation} 
where ${\bf 1}_U$ denotes the indicator function of the set $U$.
\end{itemize}
The existence of a Liapunov function is the main technical result of
this paper (see Section \ref{dissipation}) and the condition {\bf H1}
is crucial to obtain it. Note that the time derivative of the
(averaged) energy
\bd
\frac{d}{dt} \bE_x[G(x(t))] \,=\, \gamma \bE_x [\rm Tr(T) - r^2(t)]\,,
\ed
is not necessarily negative. But it is the case, as follows from our
analysis below that, that for $t>0$, $\bE_x[G(x(t)) -G(x)] < -c
G(x)^{2/k_2}$ for $x$ sufficiently large.

A nice interpretation of a Liapunov bound of the form \eqref{ll} is in
terms of hitting times. Let $\tau_U$ denote the first time the
diffusion $x(t)$ hits the set $U$; then Eq. \eqref{ll} implies that
$\tau_U$ is exponentially bounded. We will show that for any $a>0$,
{\em no matter how large}, we can find a compact set $U=U(a)$ such
that
\bd
\bE_x[ e^{a \tau_U}] \, < \, \infty   \,,
\ed
for all $x \in X$. So except for
exponentially rare excursions the Markov process $x(t)$ lives on the
compact set $U$. Combined with the fact that the process has a smooth
law, this provides an intuitive picture of the exponential convergence 
result of Theorem \ref{main}.

\section{Liapunov Function and Hitting Times} \label{dissipation}

\subsection{Scaling and Deterministic Energy Dissipation}

We first consider the question of energy dissipation for the following
deterministic equations
\begin{eqnarray}
{\dot q} \,&=&\, p \,,  \nn \\
{\dot p} \,&=&\, - \nabla_{q} V(q) - \Lambda^T r \,, \nn\\
{\dot r} \,&=&\, - \gamma r + \Lambda p   \,,\label{e30}
\end{eqnarray} 
obtained from Eq.\eqref{e23} by setting $T_1=T_n=0$, corresponding an
initial condition of the reservoirs with energy $0$.  A simple
computation shows that the energy $G(p,q,r)$ is non-increasing along
the flow $x(t)=(p(t),q(t),r(t))$ given by Eq.\eqref{e30}:
\bd
\frac{d}{dt} G(p(t),q(t),r(t)) \,=\, - \gamma r^2(t)  \, \le \, 0\,.
\ed
We now show by a scaling argument that for any initial condition 
with sufficiently high energy, after a small time, a substantial 
amount of energy is dissipated.

At high energy, the two-body interaction $U^{(2)}$ in the potential 
dominates the term $U^{(1)}$ since $k_2 \ge k_1$ and so for an 
initial condition with energy $G(x)=E$, the natural time scale -- 
essentially the period of a single one-dimensional oscillator in the potential 
$|q|^{k_2}$ -- is $E^{1/k_2 - 1/2}$.
We  scale a solution of Eq.\eqref{e30} with initial energy $E$
as follows
\begin{eqnarray}
\tp(t) \,&=&\, E^{-\frac{1}{2}}p( E^{\frac{1}{k_2} - \frac{1}{2}} t)\,, 
\nn \\
\tq(t) \,&=&\, E^{-\frac{1}{k_2}}q( E^{\frac{1}{k_2} - \frac{1}{2}} t)\,, 
\nn \\
\tr(t) \,&=&\, E^{-\frac{1}{k_2}}r(E^{\frac{1}{k_2} - \frac{1}{2}}  t)\,. 
\label{e31}  
\end{eqnarray}
Accordingly the energy scales as
$G(p,q,r)=E \tG_E(\tp,\tq,\tr)$,
where
\begin{eqnarray}
\tG_E(\tp,\tq,\tr)\,&=&\,  E^{\frac{2}{k_2}-1} \frac{\tr^2}{2} + 
\frac{\tp^2}{2} + \tV_E(\tq) \,, \nonumber \\
\tV_E(\tq)\,&=&\, \sum_{i=1}^n 
\tU^{(1)}(\tq_i) + \sum_{i=1}^{n-1} \tU^{(2)}(\tq_i - \tq_{i+1}) \,, \nn\\
\tU^{(i)}(\tx) \,&=&\, E^{-1} \tU^{(i)} (E^{\frac{1}{k_2}} x )\,, \quad
i=1,2\,. \nn
\end{eqnarray}
The equations of motion for the rescaled variables are 
\begin{eqnarray}
{\dot \tq } \,&=&\, \tp \,,  \nn \\
{\dot \tp } \,&=&\, - \nabla_{\tq} \tV_E(\tq) - E^{\frac{2}{k_2}-1} 
\Lambda^T r\,, \nn \\
{\dot \tr} \,&=&\, - E^{\frac{1}{k_2} - \frac{1}{2}} \gamma \tr + 
\Lambda \tp  \label{e37} \,. 
\end{eqnarray}
By assumption {\bf H1}, as $E \to \infty$ the rescaled energy becomes  
\begin{eqnarray}
&&\tG_\infty(\tp,\tq,\tr)\,\equiv\, \lim_{E\to\infty} \tG_E(\tp,\tq,\tr) 
\nn \\
&& \,=\,
\left\{ \begin{array}{lc} \tp^2/2 + \tV_\infty(\tq) & k_1=k_2 >2 {\rm 
~or~} 
k_2 >k_1 \ge 2  \\
\tr^2/2 + \tp^2/2 + \tV_\infty(\tq)  & k_1=k_2=2 \label{ginfty}
\end{array} 
\right. \,, \nn
\end{eqnarray}
where 
\bd
V_\infty(\tq) \,=\,
\left\{ \begin{array}{lc} \sum a^{(1)}\|\tq_i\|^{k_2} + 
\sum a^{(2)} \|\tq_i - \tq_{i+1}\|^{k_2} & k_1=k_2 \ge 2 \\
\sum a^{(2)} \|\tq_i - \tq_{i+1}\|^{k_2} & k_2>k_1\ge2 \label{vinfty} 
\end{array} 
\right. \,.
\ed
The equations of motion scale in this limit to 
\begin{eqnarray}
{\dot \tq } \,&=&\, \tp \,,  \nn \\
{\dot \tp } \,&=&\, - \nabla_{\tq} \tV_\infty(\tq) \,, \nn \\
{\dot \tr} \,&=&\,   \Lambda \tp  \label{e38} \,, 
\end{eqnarray}
in the case $k_2 >2$, while they scale to 
\begin{eqnarray}
{\dot \tq } \,&=&\, \tp \,,  \nn \\
{\dot \tp } \,&=&\, - \nabla_{\tq} \tV_\infty(\tq) -\Lambda^T r  \,, 
\nn \\
{\dot \tr} \,&=&\, -\gamma r +   \Lambda \tp  \,, \label{e38b}
\end{eqnarray}
in the case $k_1=k_2=2$.

\begin{remark}{\rm The scaling for the $p$ and $q$ is natural due 
to the Hamiltonian nature of the problem, but the scaling of $r$ has a
certain amount of arbitrariness. Since $G$ is quadratic in $r$, it
might appear natural to scale $r$ with a factor $E^{-1/2}$ instead of
$E^{-1/k_2}$ as we do.  On the other hand, the very definition of $r$
as an integral of $p$ suggests that $r$ should scale as $q$, as we
have chosen.  } \end{remark}

\begin{remark}{\rm Had we supposed, instead of {\bf H1}, that $k_1 > k_2$, 
then the natural time scale at high energy would be $E^{1/k_1-
1/2}$. Scaling the variables (with $k_2$ replaced by $k_1$ would yield
the limiting Hamiltonian $ \tp^2/2 + \sum a^{(1)}\|\tq_i\|^{k_1}$,
i.e., the Hamiltonian of $n$ {\em uncoupled} oscillators. So in this
case, at high energy, essentially no energy is transmitted through the
chain. While this does not necessary preclude the existence of an
invariant measure, we expect in this case the convergence to a SNS to
be much slower.  In any case even the existence of the SNS in this
case remains an open problem.}
\end{remark}

\begin{theorem} \label{detdiss}
Given $\tau >0$ fixed there are constants $c>0$ and $E_0 < \infty$ such 
that for 
any $x$ with $G(x)=E > E_0$  and any solution $x(t)$ of Eq.\eqref{e30} 
with 
$x(0)=x$ we have the estimate, for $t_E = E^{1/k_2-1/2}\tau $,
\begin{equation}\label{ddi}
G(x(t_E)) - E \, \le - c E^{\frac{3}{k_2}- \frac{1}{2}}\,.  
\end{equation}
\end{theorem}

\begin{remark}{\rm In view of Eq. \eqref{ddi}, this shows that $r$ is at 
least typically $O(E^{1/k_2})$ on the time interval 
$[0,E^{1/k_2-1/2}\tau]$.}
\end{remark}

\proof  Given a solution of Eq.\eqref{e30} with initial condition $x$ of 
energy $G(x)=E$, we use the scaling given by Eq.\eqref{e31} and we 
obtain
\begin{equation}
G(x(t_E)) - E \,=\, 
- \gamma \int_0^{t_E} dt\, r^2(t) \,=\, 
- \gamma E^{\frac{3}{k_2}-\frac{1}{2}} \int_0^\tau dt\, \tr^2(t) \,,
\label{dissip}
\end{equation}
where $\tr(t)$ is the solution of Eq.\eqref{e37} with initial condition 
$\tx$ 
of (rescaled) energy $\tG_E(\tx)=1$. 
By Assumption {\bf  H2} we may choose $E_0$ so large that for 
$E> E_0$ the critical points of $\tG_E$ are contained in, say, 
the set $\{\tG_E \le 1/2\}$. 

For a fixed $E$ and $x$ with $G(x)=E$, we show that there is a constant 
$c_{x,E} >0$ such that 
\begin{equation}\label{cxe}
\int_0^\tau dt\, \tr^2(t) \, \ge \, c_{\tx,E} \,.
\end{equation}
The proof is by contradiction, c.f. \cite{RT}. Suppose that
$\int_0^\tau dt\, \tr^2(t)=0$, then we have $\tr(t)=0$, for all $t\in
[0,\tau]$. From the third equation in \eqref{e37} we conclude that $
\tp_1(t)=\tp_n(t)=0$ for all $t\in [0,\tau]$, and so from the first
equation in \eqref{e37} we see that $\tq_1(t)$ and $\tq_n(t)$ are
constant on $[0,\tau]$. The second equation in \eqref{e37} gives then
\bd
0\,=\, {\dot \tp_1}(t) \,=\, -\nabla_{\tq_1} \tV(\tq(t)) \,=\,  
-\nabla_{\tq_1} \tU^{(1)}(\tq_1(t)) - \nabla_{\tq_1} 
\tU^{(2)}(\tq_1(t)-\tq_2(t))\,,
\ed
together with a similar equation for ${\dot p_n}$. By our assumption
{\bf H1} the map $\nabla \tU^{(2)}$ has a right inverse $g$ locally bounded
and measurable and thus we obtain
\bd
\tq_2(t) \,=\, \tq_1(t) - g(\tU^{(1)}(\tq_1(t))) \,.
\ed 
Since $\tq_1$ is constant, this implies that $\tq_2$ is also constant
on $[0,\tau]$. Similarly we see that $\tq_{n-1}$ is constant on
$[0,\tau]$.  Using again the first equation in \eqref{e37} we obtain
now $\tp_2(t)=\tp_{n-1}(t)=0$ for all $t\in [0,\tau]$. Inductively one
concludes that $\tr=0$ implies $\tp=0$ and $\nabla_\tq \tV =0$ and
thus the initial condition $\tx$ is a critical point of $\tG_E$. This
contradicts our assumption and Eq. \eqref{cxe} follows.

Now for given $E$, the energy surface $\tG_E$ is compact.  Using the
continuity of the solutions of O.D.E with respect to initial
conditions we conclude that there is a constant $c_E > 0$ such that
\bd
\inf_{\tx \in \{\tG_E=1\}} \int_0^\tau dt\, \tr^2(t) \, \ge \, c_E \,.
\ed
Finally we investigate the dependence on $E$ of $c_E$. We note that for 
$E=\infty$, $\tG_\infty$ has a well-defined limit given by 
Eq.\eqref{ginfty} and the 
rescaled equations of motion, in the limit $E\to \infty$, are given by 
Eqs. \eqref{e38}
in the case $k_2 >2$ and by  Eq. \eqref{e38b} in the case $k_1=k_2=2$. 
Except in the case $k_1=k_2=2$ the energy surface $\{\tG_\infty=1\}$ is 
{\em
not} compact. However, in the case $k_1=k_2>2$, the Hamiltonian
$\tG_\infty$ and the equation of motion are invariant under the
translation $r \mapsto r +a$, for any $a \in \real^{2d}$. And in
the case $k_2>k_1>2$ the Hamiltonian $\tG_\infty$ and the equation of
motion are invariant under the translation $r \mapsto r +a$ $q
\mapsto q + b$, for any $a \in \real^{2d}$ and $b \in \real^{dn}$.
The quotient of the energy surface $\{\tG_\infty=1\}$ by these 
translation, is compact.

Note that for a given $\tx \in \{\tG_\infty=1\}$ a similar argument as
above show that $\int_0^\tau dt (\tr+a)^2 > 0$, for any $a>0$ and
since this integral clearly goes to $\infty$ as $a \rightarrow \infty$
there exists a constant $c_{\infty}> 0$ such that
\bd
\inf_{\tx \in \{\tG_\infty = 1\}} \int_0^\tau \tr^2(t) \, dt > 
c_{\infty}\,.
\ed
Using again that the solution of O.D.E depends smoothly on its parameters, 
we 
obtain
 \bd
\inf_{E > E_0} \inf_{\tx \in \{\tG_E = 1\}} 
\int_0^\tau dt\, \tr^2(t) \,>\, c \,.
\ed
This estimate, together with Eq. \eqref{dissip} gives the conclusion of 
Theorem \ref{detdiss}. \qed 

\subsection{Approximate Deterministic Behavior of Random Paths}
In this section we show, that at sufficiently high energies, the
overwhelming majority of the random paths $x(t)=x(t,\omega)$ solving
Eqs.\eqref{e23} follows very closely the deterministic paths $x_{\rm
det}$ solving Eqs.\eqref{e30}. As a consequence, for most random paths
the same amount of energy is dissipated into the reservoirs as for the
corresponding deterministic ones.  We need the following {\em a
priori} ``no-runaway'' bound on the growth of $G (x(t))$.

\begin{lemma} \label{apriori} Let $\theta \le ( \max \{ T_1, T_n\})^{-1}$. Then 
$\bE_x[\exp{(\theta G(x(t)))}]$ is well-defined and satisfies the bound
\begin{equation}\label{qw1}
\bE_x[\exp{(\theta G(x(t)))}] \, \le \, \exp{(\gamma {\rm Tr}(T) \theta t)} 
\exp{ (\theta G(x))}\,.
\end{equation} 
Moreover for any $x$ with $G(x)=E$ and any $\delta >0$ we have the estimate 
\begin{equation}\label{qw2}
\bP_x \left\{ \sup_{0\le s\le t} G(x(s)) \ge (1+\delta) E \right\} \, \le \, 
\exp{(\gamma {\rm Tr}(T) \theta t)} \exp{(-\delta \theta E)} \,.
\end{equation}
\end{lemma}

\begin{remark}{\rm The lemma shows that for $E$ sufficiently large,  
with very high probability, $G(x(t))=O(E)$ if $G(x)=E$. The assumption on 
$\theta$ here arises naturally in the proof, where we need $(1-\theta 
T)\ge0$, cf. Eq. \eqref{llap}.
}
\end{remark}

\proof For $\theta \le (\max \{ T_1, T_n\})^{-1}$  we have the bound 
(the generator $L$ is given by Eq. \eqref{generator})
\begin{eqnarray}
L \exp{(\theta G(x))}
\,&=&\, \gamma \theta \exp{(\theta G(x))} \left( {\rm Tr}(T) - r (1-\theta T) r 
\right)\nn \\
\,&\le&\, \gamma \theta {\rm Tr}(T) \exp{(\theta G(x))}  \,, \label{llap}
\end{eqnarray}
so that for the function $W(t,x) = \exp{(-\gamma \theta {\rm Tr}(T)t)}
\exp{(\theta G(x))}$ we have the inequality $(\partial_t+L ) W(t,x)
\le 0$.  We denote $\sigma_R$ as the exit time from the set $\{G(x) < R\}$,
i.e., $\sigma_R = \inf\{ t \ge 0, G(x(t)) \ge R \}$. If the initial
condition $x$ satisfies $G(x)=E<R$, we denote $x_R(t)$ the process which
is stopped when it exits $\{G(x) < R\}$, i.e., $x_R(t) =x(t)$ for 
$t < \sigma_R$ and $x_R(t) = x(\sigma_R)$ for $t\ge \sigma_R$. 
We set $\sigma_R(t) = \min
\{\sigma_R, t\}$ and applying Ito's formula with stopping time to the
function $W(t,x)$ we obtain
\bd
\bE_x \left[ \exp{(\theta G(x (\sigma_R(t))))} \exp{(-\gamma \theta
{\rm Tr}(T) \sigma_R(t))} \right] - \exp{(\theta G(x))} \le 0 \,,
\ed
thus 
\begin{equation}\label{kli}
\bE_x \left[ \exp{(\theta G(x (\sigma_R(t))))} \right] \,\le \, 
 \exp{(\gamma \theta {\rm Tr}(T) t)}  \exp{(\theta G(x))} \,.
\end{equation}
Since 
\begin{eqnarray}
\bE_x \left[ \exp{(\theta G(x (\sigma_R(t))))} \right] \,&\ge&\, 
\bE_x \left[ \exp{(\theta G(x (\sigma_R(t))))} {\bf 1}_{\sigma_R < t} \right] \nn \\ 
\,&=&\, \bP_x \{\sigma_R < t \} \exp{(\theta R)}\,, \nn
\end{eqnarray}
we obtain the bound 
\bd
\bP_x \{\sigma_R < t \}\, \le  \exp{(\gamma \theta {\rm Tr}(T) t)} 
\exp{(\theta (E - R))}\,. 
\ed
As a consequence $\bP_x \{\sigma_R < t \} \rightarrow 0$  as 
$R \rightarrow \infty$ and thus 
the Markov process $x(t)$ is non-explosive. 

It follows that $G(x_R(t)) \rightarrow G(x(t))$ almost surely as 
$R \rightarrow \infty$, so by the Fatou lemma we obtain from Eq. 
\eqref{kli} the bound Eq. \eqref{qw1}. The bound Eq. \eqref{qw2} is 
obtained by noting that the left side is equal to
\bd
\bP_x \{ \sigma_{E(1+\delta)} < t\} \,\le\, \exp{(\gamma \theta {\rm Tr}(T) t)} 
\exp{(-\delta \theta  E)}\,,
\ed
and this concludes the proof of Lemma \ref{apriori}.

We have the following ``tracking'' estimates to the effect that the
random path closely follows the deterministic one at least up to time
$t_E$ for a set of paths which have nearly full
measure.  We set $\Delta x (t)\equiv x (t,\omega)-x_{\det}(t) =
(\Delta r(t), \Delta p(t), \Delta q(t))$ with both $x(t)$ and
$x_{\det}(t)$ having initial condition $x$. Let 
\bd
S(x,E,t)\,=\, \{ x(\cdot)\,;\, G(x)=E {\rm ~and~} \sup_{0\le s\le t} G(x(s)) <2E \}\,.
\ed
By Lemma \ref{apriori}, $\bP \{ S(x,E,t)\} \ge 1 - \exp{(\gamma\theta {\rm Tr}(T)t - \theta E)}$.

\begin{prop} \label{tracking} 
There exist constants $E_0 < \infty$ and  $c>0$ such that for paths $x(t,\omega) \in S(x,E,t_E)$
with $t_E =E^{1/k_2/-1/2}\tau$ and $E > E_0$ we have 
\begin{equation}\label{ttr}
\sup_{0 \le t \le t_E} \left(\begin{array}{c} 
\| \Delta q(t) \|\\ \| \Delta p(t) \| \\ \|\Delta r(t)\| 
\end{array}\right)
\,\leq\,  c \sup_{0\leq t \leq t_E} \|\sqrt{2\gamma T}\omega (t)\| 
\left(\begin{array}{c} 
E^{\frac{2}{k_2}-1} \\E^{\frac{1}{k_2}-\frac{1}{2}} \\ 1 
\end{array}\right).  
\end{equation}
\end{prop}

\proof We write differential equations for $\Delta x(t)$ again
assuming both the random and deterministic paths start at the same
point $x$ with energy $G(x)=E$. These equations can be written in the
somewhat symbolic form:
\begin{eqnarray}\label{dDeltax}
d\Delta q&=& \Delta p dt \,,\nonumber\\
d\Delta p&=& \left(O(E^{1-2/k_{2}})\Delta q - \Lambda^T \Delta r \right)dt 
\,,\nonumber\\
d\Delta r&=& \left( - \gamma\Delta r + \Lambda \Delta p \right) dt+ 
\sqrt{2\gamma T}d\omega \label{deltx} 
\end{eqnarray} 
The $O(E^{1-2/k_{2}})$ coefficient refers to the difference between
forces, $-\nabla_q V(\cdot)$ evaluated at $x(t)$ and $x_{\det} (t)$;
we have that $G (x (t))\leq 2E$, so that $\nabla_qV(q)-\nabla_q
V(q_{\det})= O(\partial^2V)\Delta q= O (E^{1-2/k_2})\Delta q$.  For
later purposes we pick a constant $c'$ so large that
\bd 
\rho= \rho(x)= c'E^{1-\frac{2}{k_2}}
\,\ge\, \sup_i\sum_j\sup_{\{q: V (q)\leq
2E\}}\left|\frac{\partial^2V(q)}{\partial q_{i}\partial q_{j}}\right|
\ed 
for all sufficiently large $E$. 

In order to estimate the solutions of Eqs. \eqref{deltx}, we
consider the $3\times 3$ matrix which bounds the coefficients in this 
system, and which is given by 
\begin{equation} \label{matrix}
M\,=\, \left( \begin{array}{ccc} 0 & 1 & 0 \\ \rho & 0 & \lambda \\ 
0 & \lambda & \gamma \end{array} \right)
\end{equation}
We have the following estimate on powers of $M$; 
For $\Delta X^{(0)}= (0,0,1)^T$, we set 
$\Delta X^{(m)}\equiv M^m \Delta X^{(0)}$. For
$\alpha=\max(1,\gamma + \lambda)$, we obtain
$\Delta X^{(1)}\le \alpha (0,1,1)^T$ 
$\Delta X^{(2)}\le \alpha^2 (1,1,1)^T$ 
and for $m\geq 3$,
\begin{eqnarray} 
\Delta X^{(m)}&\,\equiv\, &
\left(\begin{array}{c} u^{(m)}\\v^{(m)}\\w^{(m)} \end{array} \right)
\,\leq \, \alpha^m 2^{m-2} 
\left(\begin{array}{c} 
\rho^{\frac{m-2}{2}}\\ \rho^{\frac{m-1}{2}}\\ \rho^{\frac{m-2}{2}}\\ 
\end{array}\right)\,, \nn
\end{eqnarray} 
where the inequalities are componentwise.
From this we obtain the bound
\begin{equation}
e^{tM} \left( \begin{array}{c} 0 \\ 0 \\ 1 \end{array}\right) \, \le \,  
 \left( \begin{array}{c} 
\frac{1}{2}(\alpha t)^2 e^{\sqrt{\rho} 2 \alpha t} \\
\alpha t e^{\sqrt{\rho} 2 \alpha t } \\
1 + \alpha t + \frac{1}{2}(\alpha t)^2 e^{\sqrt{\rho} 2 \alpha t}
 \end{array}\right)\,. 
\label{etmbound}
\end{equation}
If $0 \le t \le t_E$ we have $\sqrt{\rho}t<\sqrt{c'}$.   
Then the  exponentials in the above equation are bounded, and 
\begin{equation}\label{Deltasup}
e^{tM} \left( \begin{array}{c} 0 \\ 0 \\ 1 \end{array}\right)  \, \leq \, 
c \left(\begin{array}{c}
1/\rho \\ 1/\sqrt{\rho}\\1 \end{array}\right) \,,
\label{etmbound2}
\end{equation}
for some constant  $c$.

Returning now to the original differential equation system
Eq.(\ref{deltx}), we write this equation in the usual integral
equation form:
\begin{eqnarray} 
\left( \begin{array}{c} \Delta q(t) \\ \Delta p(t)\\ \Delta r(t) 
\end{array}
\right) 
\,&=&\,  \int_{0}^{t}{\left(\begin{array}{c} \Delta p(s) \\ 
-\nabla_q V(q(s,\omega))\, ds + \nabla_q V(q_{\det}(s)) - \Gamma^T \Delta 
r(s) 
\\ 
-\gamma \Delta r(s) + \Lambda \Delta p(s) \end{array} \right)} \nn \\
&& \phantom{XXX} + \left( \begin{array}{c} 0 \\ 0\\ \sqrt{2\gamma T} 
\omega(t) \end{array} \right)\,. 
\label{iter1}
\end{eqnarray}
From this we obtain the bound
\bd
\left( \begin{array}{c} \|\Delta q(t)\| \\ \|\Delta p(t)\| \\ \|\Delta 
r(t)\| 
\end{array} \right) \, \le \,  \int_0^t   
M \left( \begin{array}{c} 
\|\Delta q(t)\| \\ \|\Delta p(t)\| \\ \|\Delta r(t)\| \end{array} 
\right) \,ds
+ 
\left( \begin{array}{c} 0 \\ 0\\ \omega_{\max} \end{array} \right)\,,
\ed 
where $M$ is the matrix given by 
Eq.\eqref{matrix}, and $\omega_{\max}= \sup_{t\le t_E}\|\sqrt{2 \gamma T}\omega(t)\|$. 
Note that the solution of the integral equation 
\begin{equation}
\Delta X(t)\,=\, \int_0^t ds\,  M \Delta X(s)  + \left( \begin{array}{c} 
0 \\ 0 \\  \omega_{\max} \end{array}\right) \,,
\label{iter2}
\end{equation}
is $\Delta X(t)= \exp{(tM)} (0,0,\omega_{\max})^T$.
We can solve both Eq.\eqref{iter1} and Eq.\eqref{iter2} by iteration. 
Let $\Delta x_m(s)$, $\Delta X_m(s)$
denote the respective $m^{th}$ iterates (with
$\Delta x_0(s)=(0,0, \sqrt{2\gamma T}\omega(s))^T$, and  
$\Delta X_0(s)=(0,0,\omega_{\max})^T$, $0\leq s\leq t_E$). 
The $\Delta X_m$'s are monotone increasing in $m$.
Then it is easy to see that 
\bd 
\left(\begin{array}{c} \|\Delta q_m(t)\| \\ \|\Delta p_m(t)\| \\ \|\Delta 
r_m(t)\|
\end{array} \right)
\,\le\, \Delta X_m(t) \, \le \, \Delta X(t)\,,  
\ed 
for each iterate. By Eqs.\eqref{etmbound}, \eqref{etmbound2}, and 
the definition of $\rho$ the conclusion Eq. \eqref{ttr} follows. \qed

As a consequence of Theorem \ref{detdiss} and Proposition
\ref{tracking} we obtain

\begin{corollary} \label{randiss}
Let $\Omega(E)=E^{\alpha}$ with $\alpha < 1/k_2$ and assume that $w(t)$ is 
such that  $\sup_{0\le t \le t_E} \|\sqrt{2\gamma T} \omega(t) \| \le \Omega(E)$
and $x(\cdot,\omega) \in S(x,E,t_E)$.  
Then there are constants $c>0$ and $E_0 < \infty$ such that all paths 
$x(t,w)$ with initial condition $x$ with $G(x)=E > E_0$ satisfy the bound 
\begin{equation}
\int_0^{t_E} r^{2}(s) ds \geq c E^{\frac{3}{k_2} - \frac{1}{2}}\,.
\end{equation}
\end{corollary}

\begin{remark}\label{rem39}{\rm For large energy $E$, paths {\em not} 
satisfying the hypotheses of the corollary have measure bounded by 
\begin{eqnarray}
&&\bP_x \{\sup_{0\leq s\leq t_E}\|\sqrt{2\gamma T}\omega\|> \Omega(E) \} 
+ \bP \{ S(x,E,t_E)^C \} \nn\\ 
&& \, \le \, 
\frac{a}{2} \exp\left( - \frac{\Omega(E)^2}{b \gamma T_{\max} t_E}\right) 
+ \exp{(\theta( \gamma {\rm Tr}(T) t_E -E)) } \nn\\
&& \, \le \,  
a\exp\left( - \frac{\Omega(E)^2}{b \gamma T_{\max} t_E}\right) \,,
\label{ldbound}
\end{eqnarray}
where $a$ and $b$ are constants which depend only on the dimension of 
$\omega$. Here we have used the reflection principle to estimate the first 
probability and Eq. \eqref{qw2} and the definition of $S$ to estimate the 
second probability. For $E$ large enough, the second term is small relative 
to the first. 
}

\end{remark}

\noindent {\em Proof}: It is convenient to introduce the $L^{2}$-norm
on functions on $[0,t]$, 
  $\|f\|_{t}\equiv \left(\int_0^t\|f(s)\|^{2}ds \right)^{1/2}$.
By Theorem \ref{detdiss}, there are constants $E_1$ and $c_1$ such 
that for $E > E_1$
the deterministic paths $x_{\det} (s)$ satisfy the bound
\bd
\|r_{\det}\|_{t_E}^2\,=\,\int_0^{t_E} r_{\det}^{2}(s) ds \geq c_1 
E^{\frac{3}{k_2} - \frac{1}{2}} \,.
\ed
By Proposition \ref{tracking}, there are constants $E_2$ and $c_2$ 
such 
that $\|\Delta r(s)\| \leq c_2 \Omega(E)$, 
uniformly in $s$, $0 \leq s\leq t_E$, and uniformly in $x$ with 
$G(x)>E_2$.
So we have
\bd
 \|r\|_{t_E} \geq 
\|r_{\det}\|_{t_E}-\|\Delta r\|_{t_E} 
\geq  \left(c_1 E^{\frac{3}{k_2}-\frac{1}{2}} \right)^{1/2}-c_2 
\Omega(E) \left(E^{\frac{1}{k_2}-\frac{1}{2}}\right)^{1/2}\,.
\ed
But the last term is $O(E^{\alpha -1/4+1/2k_{2}})$, which is of lower
order than the first since $\alpha <1/k_2$, so the corollary follows,
for an appropriate constant $c$ and $E$ sufficiently large.  \qed

\subsection{Liapunov Function and Exponential Hitting Times} 
With the estimates we prove now our main technical result. 

\begin{theorem} \label{liapunov}
Let $s>0$ and $\theta < \theta_0 \equiv (\max\{T_1,T_n\})^{-1}$. 
Then there are a compact set $U=U(s, \theta)$ and constants 
$\kappa=\kappa(U,s,\theta) < 1$ and $L=L(U,s,\theta) < \infty$ such that 
\begin{equation}
T^s \exp{(\theta G)}(x) \, \le \, \kappa \exp{(\theta G)}(x) + L {\bf 
1}_U(x)\,.
\label{lll}
\end{equation}
where ${\bf 1}_U$ is the indicator function of the set $U$. The constant 
$\kappa$ can be made arbitrarily small by choosing $U$ large enough.
\end{theorem}

\proof 
For any compact set U and for any $t$, $T^s \exp{(\theta G)}(x)$ is a 
bounded function, uniformly on $[0,t]$.  So, in order to prove
Eq.\eqref{lll}, we only have to prove that there exist a compact set
$U$ and $\kappa < 1$ such that
\bd
\sup_{x \in U^{\rm C}} 
\bE_x \left[ \exp{ \left( \theta ( G(x(s))  - G(x) ) \right)} \right] 
\, \le \, \kappa \, < \, 1
\,.      
\ed
Using Ito's Formula to compute $G(x(s)-G(x)$ in terms of a stochastic 
integral we obtain  
\begin{eqnarray}
&&\!\!\!\!\bE_x \left[ \exp{ \left(  \theta ( G(x(s))  - G(x)) \right)} 
\right] 
\nn \\
&&\!\!\!\!\,=\, 
\exp{(\theta \gamma \trace(T)s)} 
\bE_x \left[ \exp{\left( - \theta \int_{0}^{s} \gamma r^{2}\, dt 
+ \theta \int_{0}^{s} \sqrt{2\gamma T} r d\omega(t)  \right)} \right] \,. 
\end{eqnarray}
For any $\theta < \theta_0$, we choose $p>1$ such that 
$\theta p < \theta_0 $.  Using H\"older inequality we obtain, 
\begin{eqnarray}
&&\bE_x \left[ \exp{ \left( - \theta \int_{0}^{s} \gamma r^{2}\,dt 
+ \theta \int_{0}^{s} \sqrt{2\gamma T} r d\omega(t)  \right) } \right] 
\nn \\
&&\,=\, \bE_x \left[ \exp{\left( - \theta \int_{0}^{s} \gamma r^{2}\,dt 
+ \frac{p\theta^2}{2} \int_{0}^{s} (\sqrt{2\gamma T} r)^2\,dt  \right)} 
\times \right.
\nn \\
&&\phantom{ \,=\,\bE_x \left[ \right.} \left. 
\times \exp{\left( -\frac{p\theta^2}{2} \int_{0}^{s} 
(\sqrt{2\gamma T} r)^2 \,dt  + \theta\int_{0}^{s} \sqrt{2\gamma T} r 
d\omega(t)
\right)} \right] \nn \\
&&\,\le\, \bE_x \left[ \exp{\left( - q\theta \int_{0}^{s} \gamma r^{2}\,dt 
+ \frac{qp\theta^2}{2} \int_{0}^{s} (\sqrt{2\gamma T} r)^2\, dt  \right)} 
\right]^{1/q} \times \nn \\
&& \phantom{\,\le\,} \times 
\bE_x \left[ \exp{\left( -\frac{p^2\theta^2}{2} \int_{0}^{s} 
(\sqrt{2 \gamma T} r)^2\,dt  + \theta p \int_{0}^{s}  
\sqrt{2\gamma T}r d\omega(t) \right)} \right]^{1/p} \nn \\
&&\,=\, \bE_x \left[ \exp{\left( - q\theta \int_{0}^{s} dt\, \gamma r^{2} 
+ \frac{qp\theta^2}{2} \int_{0}^{s} dt\, (\sqrt{2\gamma T} r)^2  \right)} 
\right]^{1/q} \nn
\end{eqnarray}
Here, in the next to last line, we have used the fact that the second
factor is the expectation of a martingale (the integrand is
non-anticipating) with expectation $1$. 
Finally we obtain the bound
\begin{eqnarray} \label{sdd}
&&\bE_x \left[ \exp{  \left( \theta(G(x(s))-G(x)) \right)} \right] 
\nonumber\\
&\leq& \exp{\left( \theta \gamma \trace(T) s \right)} 
\bE_{x}\left[ \exp{\left( -q\theta (1- p\theta T_{\max}) 
\int_{0}^{s} dt\,\gamma r^{2} \right)} \right]^{1/q}
\end{eqnarray}

In order to proceed we need to distinguish two cases according if
$3/k_2 -1/2 >0$ or $3/k_2 -1/2 \le 0$ (see Corollary
\ref{randiss}). In the first case we let $E_0$ be defined by
$s=E_0^{1/k_2 -1/2}\tau$.  For $E > E_0$ we
break the expectation Eq.  \eqref{sdd} into two parts according to 
whether the paths satisfy the hypotheses of Corollary \ref{randiss} or 
not. 
 For the first part we use Corollary
\ref{randiss} and that $\int_0^{s} r^2(s) ds \ge \int_0^{t_E} r^2(s)
\ge cE^{3/k_2-1/2}$; for the second part we use estimate
\eqref{ldbound} in Remark \ref{rem39} on the probability of unlikely
paths together with the fact that the exponential under the expectation in
Eq. \eqref{sdd} is bounded by $1$.  We obtain for all $x$ with $G(x)=
E >E_0$ the bound
\begin{eqnarray} \label{fty}
&& \bE_x \left[ \exp{  \left( \theta(G(x(s))-G(x)) \right)} \right]  
\leq \exp{\left( \theta \gamma \trace(T) t_{E_0} \right)}  \times 
\nonumber \\
&& \quad \times  \left[  
\exp{\left( -q\theta (1- p\theta T_{\max}) c E^{\frac{3}{k_2} 
-\frac{1}{2}}\right)}  
 +  a \exp{ \left( - \frac{ \Omega(E)^2 \theta_0 }{ b \gamma t_E 
}\right)} 
\right]^{1/q} \,. 
\end{eqnarray}
Choosing the set $U = \{x\,;\, G(x) \le E_1 \}$ with $E_1$  large enough 
we can make the term in Eq. \eqref{fty} as small as we want.  

If $3/k_2 -1/2 \le 0$, for a given $s$ and a given $x$ with $G(x)=E$
we split the time interval $[0,s]$ into $E^{1/2-1/k_2}$ pieces
$[t_j,t_{j+1}]$, each one of size of order $E^{1/k_2-1/2}s$. For the
``good'' paths, i.e., for the paths $x(t)$ which satisfy the hypotheses
of Corollary \ref{randiss} on each time interval $[t_j,t_{j+1}]$, the tracking
estimates of Proposition \ref{tracking} imply that $G(x(t))=O(E)$ for
$t$ in each interval. Applying Corollary \ref{randiss} and using that
$G(x(t_j))=O(E)$ we conclude that $\int_0^s r^2(s) ds$ is at least of
order $E^{3/k_2-1/2}\times E^{1/2 -1/k_2} = E^{2/k_2}$. The
probability of the remaining paths can be estimated, 
using Eq. \eqref{ldbound}, not to exceed
\bd
1 - \left( 1-   a \exp{ \left( - \frac{ \Omega_{\max}^2 \theta_0 }{ b 
\gamma t_E 
}\right)} \right)^{E^{\frac{1}{2}-\frac{1}{k_2}}} \,.
\ed
The remainder of the argument is essentially as above, Eq. \eqref{fty} and 
this concludes 
the proof of Theorem \ref{liapunov}. \qed

The existence of the Liapunov function given by Eq.\eqref{lll} can be
interpreted in terms of hitting times.  Let $\tau_{U}$ be the time for
the diffusion $x (t)$ to hit the set $U$.

\begin{theorem} \label{hitting} Assume that $\theta < (\max\{T_1, 
T_n\})^{-1}$. For any (arbitrarily large) $a > 0$ there exist a constant 
$E_0=E_0(a) > 0$ such that for $U=\{ x\,;\, G(x)\le E_0\}$ and $x \in U^C$ we have 
\begin{equation}\label{expbound}
\bE_x \left[ e^{ a \tau_U} \right] \, < \, e^a + (e^a-1) 
\exp{(\theta (G(x)- E_0))} \,.
\end{equation}
\end{theorem}

\proof Let $s=1$ and $\theta < \theta_0$ be given, we set
$\kappa=\exp{(-a)}/2$ and take $U$ to be the set given by Thm. \ref{liapunov}. 
Let $X_n$ be the Markov chain defined by $X_n= x(n)$ and $N_{U}$  be the
least integer such that $X_{N_{U}}\in U$. Then 
\begin{eqnarray}\label{m1}
\bE_{x}[e^{a \tau_U}] &\,\leq\,& \bE_{x}[e^{ a N_{U}}]\,,
\end{eqnarray}
so that to estimate the exponential hitting time, it
suffices to estimate the exponential ``step number''.

Using Chernov's inequality we obtain
\begin{eqnarray} \label{m2}
\bP_{x}\{N_{U}>n \}&= &
\bP_{x}\{-\sum_{j=1}^n(G(X_j)-G (X_{j-1})<G (x)-E_{0}, X_{j}\in U^{c}\} \nonumber\\
&\leq & e^{\theta(G(x)-E_0)} \bE_{x} \left[\prod_{j=1}^n 
e^{\theta(G(X_j)-G(X_{j-1}))},X_j\in U^{c}\right] \nonumber\\
&\leq & e^{\theta(G(x)-E_0)} \bE_{x} \left[\prod_{j=1}^{n-1} 
e^{\theta(G(X_j)-G(X_{j-1}))}\right.\nonumber\\  
&&\left. \phantom{\prod_{j=1}^{n-1}} \bE_{X_{n-1}} \left[e^{\theta(G(X_n)-G 
(X_{{n-1}})}\right],\,\,X_j\in U^{c}\right]\nonumber\\ 
&\leq& e^{\theta(G(x)-E_0)} \sup_{y\in U^{c}} \bE_{y}[e^{\theta ( G 
(X_{1})-G (y))}]\times\nonumber\\ 
&&\phantom{XXXXXXXX} \bE_{x}\left[\prod_{j=1}^{n-1} 
e^{\theta(G(X_j)-G(X_{j-1}))},X_j\in U^{c}\right]\nonumber\\
&\leq&\cdots\leq  e^{\theta(G(x)-E_0)}\left( \sup_{y\in U^{c}} 
\bE_{y}[e^{\theta ( G (X_{1})-G (y))}]\right)^n. \nn
\end{eqnarray}
By Thm. \ref{liapunov} we have
\bd
  \sup_{x\in U^{c}} \bE_{x}[e^{\theta ( G (X_{1})-G (x))}]\,<\, \kappa \,,
\ed 
and therefore we have geometric decay of 
$P_{>n}\equiv \bP_{x}\{N_{U}>n \}$ in $n$, 
$P_{>n} \le \kappa^n \exp{(\theta G(x)-E_0)}$. 
Summing by parts we obtain 
\begin{eqnarray}
&&\bE_x\left[e^{a N_U}\right]\,=\, \sum_{n=1}^\infty e^{an}
 \bP_x\{\tau_U =n\} \nn\\
&&\,\,\,=\, \lim_{M\rightarrow \infty} \left[ \sum_{n=1}^M 
 P_{>n} (e^{a(n+1)} -e^{an})  + e^a P_{>0} - e^{a(M+1)} P_{>M} \right] \nn
\end{eqnarray}
which, together with Eq. \eqref{m1} gives Eq. \eqref{expbound}. \qed

\section{Accessibility and Strong Feller Property}\label{controlandfeller}

In this section we prove that the Markov process is strong Feller and 
moreover we show that it is strongly aperiodic in the sense that for all 
$t>0$, all $x \in X$ and all open sets $A \subset X$ we have $P_t(x,A) >0$. 
Both results imply immediately that $x(t)$ has at most one invariant 
measure: Since the process is strong Feller the invariant measure (if it 
exists) has a smooth density which is everywhere positive by the property 
of aperiodicity. Obviously no two different such measures can exist. 

The strong Feller property is an immediate consequence of the hypoelliptic 
properties of the generator $L$ of the diffusion. The result is an easy 
consequence of the estimates in \cite{EPR2,EH}, since there much stronger global 
hypoelliptic estimates are proven (under stronger conditions on the 
potential $U^{(2)}$). We present here the argument for completeness. 

The generator of the Markov process $x(t)$ can be written 
in the form
\bd
L\,=\, \sum_{i=1}^{2d} X_i^2 + X_0\,. 
\ed 
If the Lie algebra generated by the set of commutators
\begin{equation}\label{rank}
\{ X_i\}_{i=1}^{2d}\,, \quad  \{ [X_{i},X_{i}] \}_{i,j =0}^{2d}\,, \quad
\{ [[X_{i},X_{j}],X_{k}] \}_ {i,j,k =0}^{2d}\,, \quad \cdots  
\end{equation}
has rank ${\rm dim}(X)$ at every point $x\in X$, then the Markov process
has a $\calC^\infty$ law. In particular it is strong Feller. This is a
consequence of H\"ormander Theorem \cite{Ho,Ku} or it can be proved
directly using Malliavin Calculus developed by Malliavin, Bismut,
Stroock and others (see e.g. \cite{No}).

\begin{prop} If {\bf H2} holds then the generator $L$ given by 
Eq.~\eqref{generator} satisfies the rank condition \eqref{rank}. 
\end{prop}

\proof This is a straightforward computation. The vector fields $X_i$, 
$i=1,\cdots 2d$ gives 
$\partial_{r_i^{(j)}}$, $i=1,n$,  $j=1,\cdots,d$. The commutators 
\begin{eqnarray}
\left[ \partial_{r_1^{(j)}}\,,\, X_0 \right] & = & \gamma 
\partial_{r_1^{(j)}} -
\lambda \partial_{p_1^{(j)}} \,, \nn \\ 
\left[ \left[ \partial_{r_1^{(j)}} \,,\, X_0 \right] \,,\, X_0 \right] &=& 
\gamma^2 \partial_{r_1^{(j)}} - 
\gamma \lambda \partial_{p_1^{(j)}} - \lambda \partial_{q_1^{(j)}} \,,\nn
\end{eqnarray}
yield the vector fields $\partial_{p_1^{(j)}}$ and $\partial_{q_1^{(j)}}$. 
Further
\bd
\left[ \partial_{q_1^{(j)}} \,,\, X_0\right] \,=\, \sum_{l=1}^d 
\frac{\partial^2 V}{\partial_{q_1^{(j)}} \partial_{q_1^{(l)}}}(q)  
\partial_{p_1^{(l)}} 
+  \sum_{l=1}^d
\frac{\partial^2 U^{(2)}}{\partial_{q_1^{(j)}} 
\partial_{q_2^{(l)}}}(q_1-q_2)  
\partial_{p_2^{(l)}} \,.
\ed
If $U^{(2)}$ is strictly convex, this yields  $\partial_{p_2^{(j)}}$ while 
in the general 
case we need to consider further the commutators 
\begin{eqnarray}
&& \left[ \partial_{q_1^{(j_1)}}\,,\, \left[ \cdots\,,\,
\left[\partial_{q_1^{(j_{m-1})}} \,,\, \sum_{l=1}^d \frac{\partial^2
U^{(2)}}{\partial_{q_1^{(j_m)}} \partial_{q_2^{(l)}}}(q_1-q_2)
\partial_{p_2^{(l)}} \right] \right] \right] \nn \\
&&\quad \, = \,  \sum_{l=1}^d
\frac{\partial^{m+1} U^{(2)}} {\partial_{q_1^{(j_1)}} \cdots
\partial_{q_1^{(j_m)}} \partial_{q_1^{(l)}} }(q_1-q_2)
\partial_{p_2^{(l)}} \,. \nn
\end{eqnarray} 
The condition {\bf H3} means that we can write $\partial_{p_2^{(j)}}$
as a linear combination of these commutators for every $x \in X$. The
other basis elements of the tangent space are obtained inductively
following the same procedure. \qed

We now prove the strong aperiodicity of the process $x(t)$.  This is
based on the support theorem of Stroock and Varadhan \cite{SV}.  The
support of the diffusion process $x(t)$ with initial condition $x$ on
the time interval $[0,t]$, is by definition the smallest closed subset
$S_{x,t}$ of $\calC([0,t])$ such that ${\bf P}_x [ x(t,\omega) \in
S_{x,t}]=1$.  The support can be studied using the associated control
system, i.e., the ordinary differential equation where the white noise
${\dot \omega}(t)$ is replaced by a control $u(t) \in L^1([0,T])$: For
our problem we have the control system
\begin{eqnarray}
{\dot q} \,&=&\, p \,,\nn \\
{\dot p} \,&=&\, -\nabla_q V + \Lambda^T r \,, \nn \\
{\dot r} \,&=&\, (-\gamma r + \Lambda p) + u \,, 
\label{e42}
\end{eqnarray}
and we denote $x_u(t)$ the solution of this control system with initial 
condition $x$ and control $u$.
The support theorem asserts that the support of the diffusion $S_{x,t}$ is the 
closure of the set $\{x_u\,;\, u \in  L^1([0,t])\}$. As a consequence 
${\rm supp}\, P_t(x,\cdot)$, the support of the transition probabilities is equal 
to the closure of the set of accessible points $\{ y\,;\, \exists u \in L^1([0,t]) {\rm 
~s.t.~} x_u(t)=y \}$. 

\begin{prop} If condition {\bf H1} holds then for all $t>0$, all 
$x\in X$ 
\begin{equation} 
{\rm supp}\, P_t(x,\cdot) \,=\, X\,.
\end{equation}
\end{prop}

\proof This result is proved in \cite{EPR2} under the additional condition 
that the interaction potential $U^{(2)}$ is strictly convex, in particular
$\nabla U^{(2)}$ is a diffeomorphism. Our condition {\bf H1} implies that 
$ \nabla U^{(2)}$ is surjective. We can choose an inverse 
$g : {\bf R}^d \rightarrow {\bf R}^d $ which is locally bounded. From this 
point the proof proceeds exactly as in Theorem 3.2 of \cite{EPR2} and we 
will not repeat it here.

\section{Proof of Theorem \ref{main}} \label{proof}
The proof of Theorem \ref{main} is a consequence of the theory linking
the ergodic properties of Markov process with existence of Liapunov
functions, a theory which has been developed over the past twenty
years.  The proof of these ergodic properties relies on the intuition
that the compact set $U$ together with a Liapunov function plays much
the same role as an atom in, say, a countable state space Markov
chain.  The technical device to implement this idea was invented in
\cite{AM,Nu}, and is called {\em splitting} It consists in constructing a
new Markov chain with state space $X_0 \cup X_1$, where $X_i$ are two
copies of the original state space $X$. The new chain possesses an
atom and has a projection being the original chain. The ergodic
properties of a chain with an atom are then analyzed by means of {\em
renewal theory} and a {\em coupling argument} applied to the return
times to the atom.  A complete account of this theory for a discrete
time Markov process is developed in the book of Meyn and Tweedie
\cite{MeTw}, from where the result needed here is taken (Chapter 15).

For a given $s>0$ consider the discrete time Markov chain $X_j= x(js)$
with transition probabilities $P(x,dy)\equiv P_s(x,dy)$ and semigroup
$P^j\equiv T^{js}$. By the results of Section \ref{controlandfeller},
the Markov chain is strongly aperiodic, i.e., $P(x,A)>0$ for any open
set $A$ and for any $x$ and it is strong Feller.  The exponential
bound on the hitting time given in Theorem \ref{hitting} implies in
particular that ${\bf E}_x [\tau_U]$ is finite for all $x \in X$ and
thus we have an invariant measure $\mu$ (for hypoelliptic diffusions
this is established in \cite{Kl}).  By aperiodicity and the strong
Feller property, this invariant measure is unique.

The following Theorem is proved in \cite{MeTw}: 
\begin{theorem} \label{MT}
If the Markov chain $\{X_j\}$ is strong Feller and  strongly aperiodic and 
if there is a 
function $W >1$, a compact set $U$ and
$\kappa < 1$ and $L < \infty$ such that
\begin{equation}
PW(x) \, \le \, \kappa W(x) + L{\bf 1}_U(x) \,,
\label{expd}
\end{equation}
then there exist constants $r>1$ and $R< \infty$ such that, for any x, 
\bd
\sum_n r^n \| P(x,\cdot) - \mu \|_W \, \le \, RW(x)\,,
\label{expd2}
\ed
where the weighted variation norm $\| \cdot\|_W$ is defined in
Eq. \eqref{wvar}. 
\end{theorem}
By Theorem \ref{liapunov} the assumptions of Theorem 
\ref{MT}
are satisfied with $W= \exp(\theta G)$ and $\theta < (\max
\{T_1,T_n\})^{-1}$. For the semigroup
$T^t$ we note that we have the apriori estimate 
$T^t \exp(\theta G)(x) \le \exp(\gamma \theta {\rm Tr}(T)t)\exp(\theta 
G)(x)$, cf. Lemma \ref{apriori} which shows that
$T^t$ is a bounded operator on $L^\infty_\theta(X)$ defined in Eq. 
\eqref{ltheta}. Setting $t= n s +
u$ with $0\le u < s$, and using the invariance of $\mu$ one obtains
\begin{equation}
\| T^t - \mu\|_\theta \,\le\, \| T^{n\tau} - \mu \|_\theta 
\|T^s\|_\theta \, \le \, {\tilde R} 
{\tilde r}^{-t}\,,
\label{ex01}
\end{equation}
for some ${\tilde r} > 1$ and ${\tilde R} < \infty$ or equivalently
\bd
\int^\infty_0 {\tilde r}^t \| P_t(x,\cdot) - \mu \|_{\exp{(\theta G)}} 
\, \le \, {\tilde R} \exp{(\theta G(x))}\,.
\ed
As a consequence, for any $s>0$, $T^s$ has $1$ as a simple eigenvalue
and the rest of the spectrum is contained in a disk of radius
$\rho<1$.  The exponential decay of correlations in the stationary
states follows from this.

\begin{corollary} 
There exist constants $R < \infty$ and $r > 1$ such that for all $f$, $g$ with
$f^2$, $g^2 \in L^\infty_\theta(X)$, we have
\bd
\left| \int f T^tg\, d\mu - \int f\, d\mu \int g\, d\mu\right| 
 \le R \|f^2\|_\theta^{1/2}\|g^2\|_\theta^{1/2} r^{-t} \,.
\ed
\end{corollary}

\proof If $f^2 \in L^\infty_\theta$, we have $|f(x)| \le \|f^2\|_\theta^{1/2}
\exp(\theta G(x)/2)$ and similarly for $g$. 
Further if Eq.\eqref{ex01} holds with $W=\exp{(\theta 
G)}$ it also holds for $\exp{(\theta G/2)}$ and thus for some 
$R_1 < \infty$ and $r_1 > 1$ we have 
\bd
\left |T^t g(x) - \int g\,d\mu \right| \,\le\, R_1 r_1^{-t} 
\|g^2\|_\theta^{1/2} \exp{\left(\frac{\theta G(x)}{2} \right)} \,.
\ed
Therefore we obtain
\begin{eqnarray}
\left |\int f T^tg \,d\mu - \int f\,d\mu \int g\, d\mu \right| \, &\le& 
\, \int |f(x)|
\left|T^tg(x) - \int g\,d\mu \right|\, d\mu  \nn \\
\, &\le& \, \left(\int \exp{(\theta G)} d\mu\right) R_1 r_1^{-t} 
\|f^2\|_\theta^{1/2}\|g^2\|_\theta^{1/2} \,. \nn
\end{eqnarray}
To conclude we need to show that $\int \exp{(\theta G)} d\mu < \infty$. 
This follows from
Eq.\eqref{expd} which we rewrite as
\bd
\epsilon \exp{(\theta G(x))} \,\le\, \exp{(\theta G(x))} - P\exp{(\theta G(x))} + 
L{\bf 1}_U(x)\,,
\ed
with $\epsilon=1-\kappa$. From this we obtain
\begin{equation}
\epsilon \frac{1}{N}\sum_{k=1}^N \exp{(\theta G(X_k))} \le \frac{1}{N} 
\exp{(\theta G(x))} + L \frac{1}{N}
\sum_{k=1}^N {\bf 1}_U(X_k)\,.
\label{kj}
\end{equation}
By the Law of Large Numbers the r.h.s of Eq.\eqref{kj} converges to
$L\mu(U)$ which is finite, and thus $\int \exp{(\theta G)} \,d\mu$ is
finite, too. \qed

This concludes the proof of Theorem \ref{main}.

\begin{acknowledgements} 
We would like to thank Pierre Collet, Jean-Pierre Eckmann, Servet Martinez 
and Claude-Alain Pillet for their comments 
and suggestions as well as Martin Hairer for useful comments on the 
controllability issues discussed in Section \ref{controlandfeller}. 

\end{acknowledgements}

\end{document}